# Nonlinear exceptional points in an integrated acoustic-wave oscillator for longwave infrared sensing

Linbo Shao[1,*], Zichen Xi[1], Zengyu Cen[2], Joseph G. Thomas[1], Dongyao Wang[2,3], Tanmay Singh[2], Liyan Zhu[4], Honghu Liu[5], Jun Ji[1], Yu Yao[2,*], and Yizheng Zhu[1,*]

[1]Bradley Department of Electrical and Computer Engineering, Virginia Tech, Blacksburg, VA 24061, USA
[2]School of Electrical, Computer and Energy Engineering, Arizona State University, Tempe, AZ, 85281 USA
[3]Materials Science and Engineering, Fulton Schools of Engineering, Arizona State University, Tempe, AZ 85287, USA
[4]Center for Power Electronics Systems (CPES), Virginia Tech, Blacksburg, VA 24061, USA
[5]Department of Mathematics, Virginia Tech, Blacksburg, VA 24061, USA
Corresponding authors: shaolb@vt.edu (L.S.), yuyao@asu.edu (Y.Y.), yizhu1@vt.edu (Y.Z.)

**Abstract**
Exceptional points (EP) featuring enhanced responsivity and rich dynamics have attracted extensive attentions in device developments and sensing applications. However, it remains debated whether employing EP systems is beneficial in practical sensing applications. Here, we demonstrate that a nonlinear EP in our microwave-frequency acoustic-wave oscillator improves longwave infrared (LWIR) detection under practical conditions. By phase tuning the nonlinear gain, our detector can be operated at different conditions with respect to the nonlinear EP. Compared with operation away from EP, our detector at EP shows a 33-fold improvement in responsivity and an 8.75-fold extension of 3-dB bandwidth. We observe a 6-fold enhancement in signal-to-noise ratio at an input modulation frequency of 6.2 kHz. At the incident LWIR wavelength of 9.6 μm, our detector at EP exhibits a noise equivalent power (NEP) of 310 pW·Hz$^{-1/2}$ at input frequency of 10 kHz, yielding a figure of merit, product of NEP and time constant (NEP·τ), of $9.87\times10^{-3}$ pW·Hz$^{-3/2}$, a 10-fold improvement over operation away from EP. Our integrated acoustic devices offer a versatile platform for exploring noise dynamics and developing practical sensors that exploit non-Hermitian nonlinearities.

**Introduction**
Responsivity, the relationship between the output signal and input, is a key performance metric for sensors. Beyond engineering materials, structures, and sensing mechanisms, employment of exceptional points (EP) of non-Hermitian systems can significantly enhance sensors' responsivity [1-10]. A typical EP sensor includes two coupled modes, one with gain and the other with loss, or equivalently one with less loss and the other with more loss. Near EP, gain, loss, and coupling of the sensor result in spectral degeneracies, where the eigenfrequency becomes ultra-sensitive to perturbations, i.e., input signals. In the classical dynamics, the enhancement of responsivity could be theoretically infinitely high at EP. Experimental demonstrations show significant responsivity enhancement in an optical system [1], electronic systems[7], and sub-MHz to MHz-frequency electromechanical system [5,8,9], depending on how close the sensors operate to EP. Beyond sensing applications, such EP or more broadly parity-time-symmetric systems [11] also achieve directional lasing [12,13], electromagnetically induced transparency [14], efficient wireless power transfer [15], and nonreciprocal behaviors [16-21].

Beyond technological developments of EP-based sensors, whether the signal-to-noise ratio (SNR) is or could be improved in EP-based sensing has been debated for years, often compounded by the complexity of experimental and theoretical noise limits. Experimentally, the Petermann factor in a laser results in no SNR improvement in linear EP systems [22]. Meanwhile, SNR enhancement were observed in nonlinear EP systems based on electronic circuity [4,7] and quantum defect centers [23]. We note that it is challenging to experimentally distinguish the origins of such SNR enhancement – whether from the EP or from the gain, or both, as sensors leveraging oscillating or lasing show improved SNR compared to their passive counterparts [24,25]. On the other hand, by theoretical analysis and numerical simulations, some suggest no fundamental SNR enhancement of both linear [26,27] and nonlinear [28,29] EP systems with classical gains, while some suggests potential SNR enhancement relies on an optimized data processing algorithm [30]. Protocol to study EP-based sensor in the quantum region has also been theoretically proposed [31].

Reaching EP and maintaining there is another major challenge, particularly for integrated sensors, which limits the implementation of EP-based sensing in real applications. Mismatches in frequency, gain, loss, and coupling can quickly diminish the enhancement of responsivity. Previous demonstrations utilize thermal tuning [1], nano positioned scatters [3,12], fine tuning of coupling [32,33], extra unidirectional channel [6], pump power [13], tuning capacitor [5] and tuning inductor [4] to achieve a near-EP condition. Despite these efforts, it is still challenging to maintain the EP condition on integrated sensing systems, especially in a robust and scalable manner. In this work, we introduce a more robust method to approach EP by tuning the phase of the gain in the active feedback loop.

Mechanical-based longwave infrared (LWIR) detectors [34] are of particular interests to employ such nonlinear EP systems. Unlike shorter-wavelength visible and near-infrared photodetectors, whose performance have achieved already close to the shot-noise limit, room-temperature LWIR detectors are not yet close to that. Semiconductor-based LWIR photodiodes [35-38] and bolometers [39,40] typically require cryogenic temperatures to mitigate high dark currents or thermal noises, and their sensitivities drop significantly towards long wavelengths (>8 μm). Mechanical resonators show advantages in room-temperature LWIR detection [24,40-52]. The high mechanical quality ($Q$) factors and thermally isolated structures lead to high sensitivity but at the cost of slow response and challenges to accurately readout resonant frequencies, both could be improved by employing a nonlinear EP system.

In this Article, we demonstrate an acoustic-wave nonlinear EP for LWIR detection that shows practical benefits of employing EP system. We leverage the tunable phase of the gain loop to efficiently reach the EP condition. As a temperature-based LWIR detector, we show that our EP-based configuration can significantly extend its bandwidth and thus improves measured SNR at high input frequencies. At near EP, we observe a 3-dB (10-dB) bandwidth extension by factors of 8.75× (2.08×) and a responsivity enhancement exceeding 33×, compared to those when operating away from EP. We also observe an overall signal-to-noise ratio (SNR) enhancement of 6.1× for an input LWIR signal modulated at 6.2 kHz. Our experiments and analysis render practical feasibility and technical advantages of nonlinear EP sensors. Our devices provide a practical, experimental platform to explore nonlinear-EP-based sensors from real application perspective, complementing recent theoretical and simulation works on nonlinear EP sensing [28,30].

**Acoustic-wave nonlinear EP LWIR detector**

We employ coupled acoustic-wave resonators on the bulk lithium niobate (LN) substrate (**Fig. 1a**). Acoustic-wave resonators *a* and *b* are defined by the phononic crystals, which are etched grooves on surface of LN, offering high acoustic quality ($Q$) factors and small mode volume for sensing [24]. The coupling between the resonators is determined by the number of phononic crystal periods. The design of the coupled phononic crystal resonators is summarized in **Supplementary Fig. 1**. We place interdigital transducers (IDT) inside and outside the resonator. A typical passive transmission spectrum measured from the IDT pair of one resonator shows double peaks, indicating quality ($Q$) factors of 800 and 1,400 of coupled modes and a splitting (coupling strength) of 1.5 MHz (**Fig. 1b**).

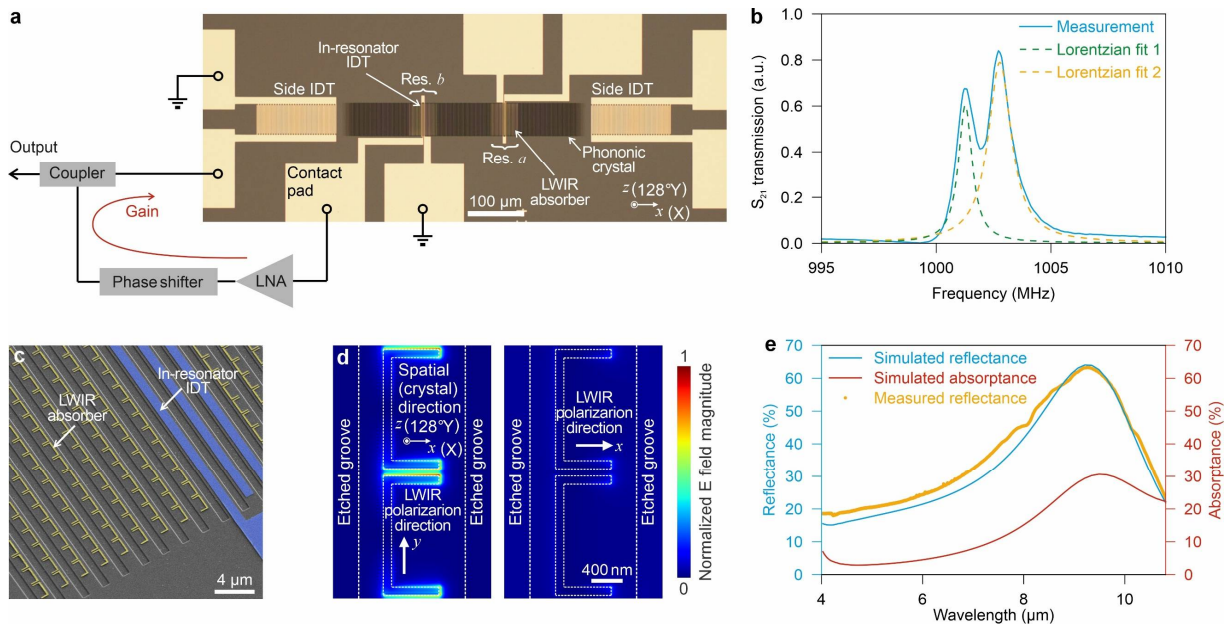

**Fig. 1| Acoustic-wave longwave infrared (LWIR) detector leveraging nonlinear exceptional-point (EP) system. a,** Configuration of our nonlinear EP LWIR detector. The acoustic wave propagates in the crystal X of LN. **b,** Transmission spectrum of the coupled resonators without gain, measured from the side interdigital transducers (IDT) to the in-resonator IDT. The transmission spectrum shows resonance at 1 GHz with a splitting of 1.5 MHz and quality ($Q$) factors of 800 and 1,400. **c,** False-colored scanning electron microscopy (SEM) image shows our phononic crystal resonator with C-shaped LWIR absorber arrays. **d,** Simulated optical electrical fields with incident polarization in device $y$ (left) and $x$ (right) directions, showing stronger resonance (absorption) for $y$ polarization. The incident LWIR wavelength is 9.6 μm. The boundary of etched grooves and metal absorbers are marked by dashed lines. **e,** Absorptance and reflectance spectra of the absorber array.

Building nonlinear-EP-based sensors, we introduce the gain to one of the resonators (resonator $b$) by forming a loop from the in-resonator IDT, passing through a low noise amplifier (LNA), an electrically controlled phase shifter, and couplers, to the side IDT. This loop provides a saturable nonlinear gain with tunable phase. In operation of our nonlinear EP detector, the gain will fully compensate for the loss, and the system starts self-oscillating (or phonon lasing). The output is outcoupled from the coupler in the gain loop, I/Q downconverted to low frequency, and captured by an oscilloscope. Experimental details are provided in **Methods**. We note that the saturable gain results in a nonlinear third-order EP [4,28].

Our acoustic-wave sensor realizes a thermal based LWIR detection: when the incident LWIR is absorbed by the metasurface absorbers, it induces temperature changes of the acoustic-wave resonators and leads to the resonant frequency shift. The C-shaped metasurface absorbers enable polarization and wavelength selective LWIR detection (**Figs. 1c-1e** and **Supplementary Fig. 2**). LN features a thermoelastic coefficient of -70 ppm/K for the used 128Y-X device configuration [53], with its low acoustic-wave loss and strong piezoelectricity, making it a promising material platform for temperature-based mechanical sensors. Overall, EP brings two benefits to our LWIR detection: (1) the observed detector responsivity near the EP is much larger than that away from EP, (2) the detector bandwidth is extended, as EP enhance the response to the differential temperature changes between two resonators but not the common-mode temperature changes from the resonators to the environment. In other words, the EP does not enhance the responsivity to the macroscopic environment temperature fluctuations, which shifts resonant frequencies of both resonators but not driving the sensor away from the EP condition.

**Acoustic nonlinear EP system**

The dynamics of our nonlinear-EP-based LWIR detector can be described by

$$\frac{d}{dt}\begin{pmatrix}a\\b\end{pmatrix} = -i\begin{pmatrix} \omega_a + \eta\,\Delta T_a - i\frac{\gamma_a}{2} & \mu \\ \mu & \omega_b + \eta\,\Delta T_b + ig_{\rm NL}(b)e^{i\theta} - i\frac{\gamma_b}{2} \end{pmatrix}\begin{pmatrix}a\\b\end{pmatrix}.$$

Here $a$ and $b$ are amplitudes corresponding to resonator $a$ and $b$ (as shown in **Fig. 1a**) with intrinsic resonant frequencies of $\omega_a$ and $\omega_b$, respectively. $\mu$ is the coupling between the two modes, and $\gamma_a$ and $\gamma_b$ are their respective total losses. $\Delta T$ describes the temperature of individual resonators (referenced to the initial thermal equilibrium), and $\eta$ is the susceptibility of the resonant frequency to the temperature. $g_{\rm NL}(b)$ is the saturable nonlinear gain, and $\theta$ is the tunable phase of the gain loop. A sufficient gain will drive the system into self-oscillation (or phonon lasing). When the system is near the nonlinear EP, the oscillating frequency $\omega_{\rm osc}$ features enhanced responsivity to the temperature difference between two resonators, $\Delta T_a - \Delta T_b$, caused by, for example, the incident LWIR light. We note that the oscillating frequency $\omega_{\rm osc}$ corresponds to one of the eigenmodes of the system: in the weak coupling case, the oscillating frequency corresponds to the only one real eigenfrequency; in the strong coupling case, the oscillating frequency corresponds to the mode that is mathematically stable and requires lower gain. A detailed analytical analysis of the nonlinear EP system is provided in **Supplementary Note 1**. Thermodynamics and noises model of our sensor is discussed in **Supplementary Note 2**, and numerical simulation results are presented in **Supplementary Note 3**.

For a system with purely real gain, i.e. $\theta = 0$, the nonlinear EP requires two conditions: (1) the two involved modes have the same resonant frequency, $\omega_a = \omega_b$, and (2) the loss matching the coupling rate, $\gamma_a = 2\mu$. However, due to fabrication and component variances, mismatches are inevitable and will quickly diminish the enhancement of the responsivity. As our acoustic-wave devices features a $Q$ factor over 1,000, the EP-based detectors are more sensitivity to the mismatch in resonant frequency than the coupling and loss rates. We reliably extract the device parameters of the fabricated coupled resonators from the transmission spectra or displacement field measurements using our in-house optical vibrometer [54]. We note that the loss/coupling mismatch, $\frac{2\mu}{\gamma}$, can be well controlled within 3% of the EP condition, while the resonant frequency mismatch is typically on the order of linewidth, *i.e.*, $(\omega_a - \omega_b) \sim \gamma$. **Supplementary Note 4** discusses the device parameter estimations based on optical vibrometry measurements.

Instead of directly tuning the resonant frequency, here we tune our detector towards the EP by adjusting the gain phase $\theta$ (via the electrically-controlled phase shifter), which can compensate for the resonance frequency mismatch $\omega_a - \omega_b$. This approach avoids extra tunning elements (for example, microheaters) for the integrated devices. We measure the oscillating frequency $\omega_{\rm osc}$ at different gain phases (**Fig. 2**). We characterize different devices with very weak ($2\mu/\gamma \ll 1$), weak ($2\mu/\gamma < 1$), near EP ($2\mu/\gamma \sim 1$), and strong ($2\mu/\gamma > 1$) coupling rate with respect to the EP condition, corresponding to **Figs. 2a-2d**, respectively. We observe the steeper changes in oscillating frequencies, when the system is getting close to the EP condition. Notably, in the device with stronger coupling than EP condition, we observe a frequency jump. This jump corresponds to the jumps from one branch of the eigenmode to the other, as predicted by our theoretical analysis. We note that such frequency jumps have been observed in previous work [15], and we derived a full analytical prediction of such frequency jumps (see **Supplementary Note 1**). In real applications, we suggest to avoid frequency jumps by operating at slightly weak coupling region.

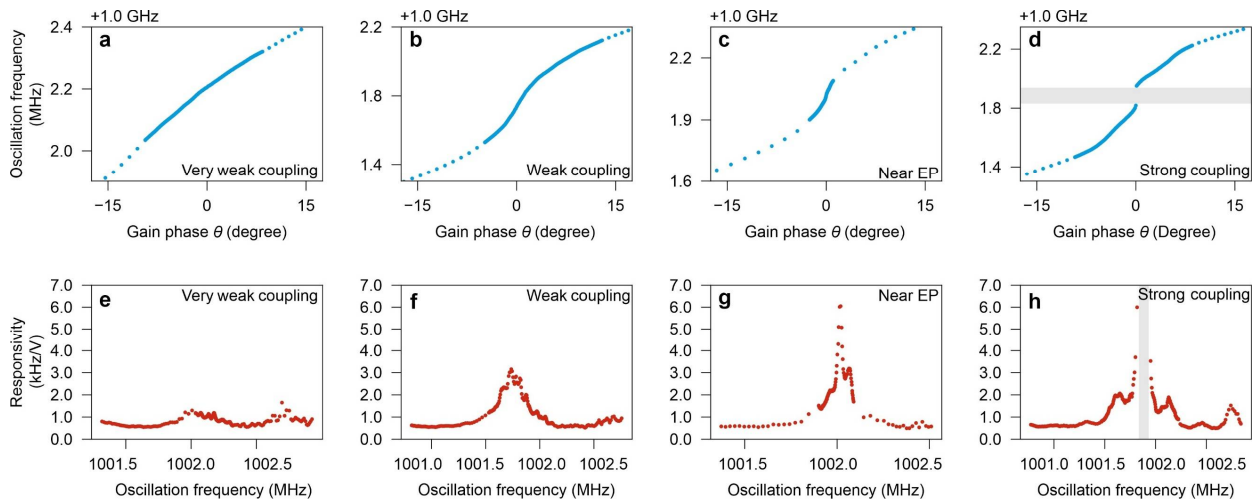

**Fig. 2 | Oscillation frequency and responsivity of our acoustic-wave systems at different coupling conditions**. **a-d.** Oscillation frequency versus external gain phase. The gain phase is relative in each plot. **e-h.** Responsivity to an electrical perturbation applied on the gain resonator. The coupling conditions are (**a, e**) very weak, (**b, f**) weak, (**c, g**) near EP and (**d, h**) strong coupling conditions. The grey region in (**d, h**) indicates the frequency jump observed under the strong coupling condition. All devices in this figure are fabricated on a same chip.

We further characterize the responsivity of our detector to an external electrical perturbation applied on the gain resonator (**Figs. 2e-2h**). The external perturbation is provided by an electrical signal through a bias-tee in the gain loop via the electro-acoustic effect [55]. We observe peak responsivity enhancement factor of 5.0, 10.5, and 9.4 for the weak, near EP, and strong coupling devices. These experimental results show that our gain phase tuning approach can effectively tune our system towards EP condition. We speculate the multiple peaks in responsivity at different frequencies could be due to the acoustic-wave and electrical interference in the gain loop.

**LWIR detection near nonlinear EP**

We demonstrate the LWIR detection using our nonlinear-EP-based detector (**Fig. 3a**); the incident LWIR at 9.6 μm wavelength is modulated by an electrical driving circuit. The modulated LWIR input modulates the resonant frequency of the phononic crystal resonator with gain, and induces the frequency modulation (FM) of the output oscillating signal. We extract the frequency of the output signal by fitting the phase change of the measured in-phase and quadrature (I/Q) signals in a short time period (**Fig. 3b** and details in **Methods**), which corresponds to desired readout bandwidth. For example, a readout bandwidth of 2 kHz means each data point of frequency is derived by a linear fitting of I/Q data of 0.5 ms. This approach has been used in our previous work on single resonator detector, showing a good SNR performance in data processing [24]. An approach equivalent to the phase fitting is also discussed in the recent theoretical work [30] to overcome certain noises related to EP.

When tuning our detector towards the EP point by adjusting the gain phase $\theta$. Here, we refer near EP condition as the gain phase $\theta$ resulting in maximum responsivity, and away from the EP as the region the responsivity does not depend on the gain phase $\theta$.

We compare the detected FM signals and their Fourier transformed spectra when the incident LWIR is modulated at different frequencies and the detector is tuned near the EP and away from the EP (**Figs. 3c-3f**). Our detector shows enhanced signal amplitude at EP as compared to that away from the EP (note the different y-axis ranges in **Figs. 3c-3f**). Signal-to-noise ratio (SNR) enhancement are clearly observed when the incident LWIR is modulated at higher frequencies (**Figs. 3d-3f**). To account for the different readout bandwidths at different modulation frequencies, the SNR is normalized to a 1-Hz bandwidth. The corresponding SNR are 40, 262, 270, and 268 $Hz^{1/2}$ at incident LWIR modulation frequency of 18, 1,920, 6,170, and 13,890 Hz when operating at near EP. In comparison, the SNR is 28, 94, 42 and <23 $Hz^{1/2}$ at same modulation frequencies when operating at away from EP. These results give direct experimental evidence that EP system can be beneficial in practical sensing applications.

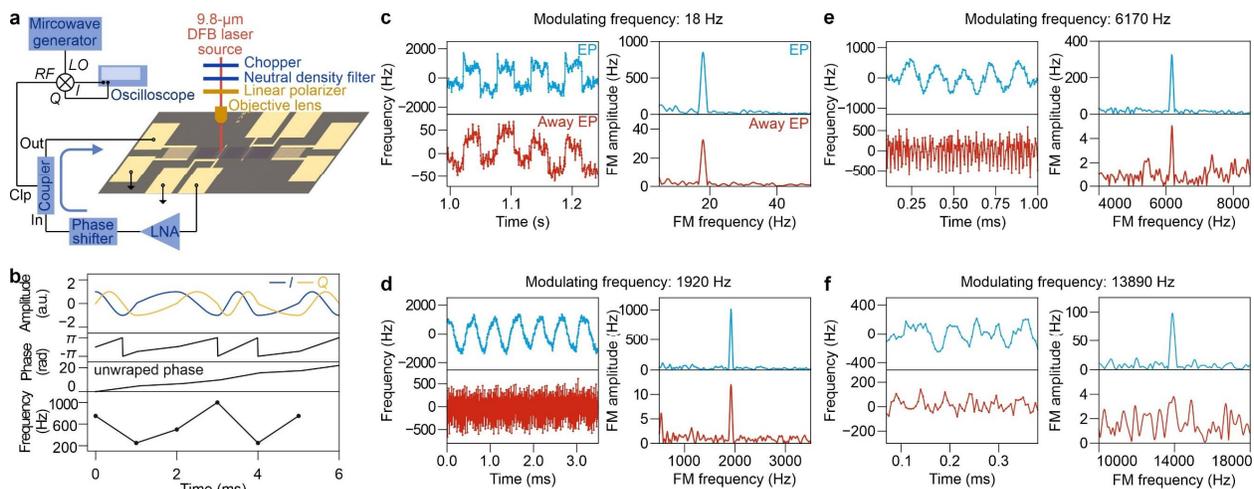

**Fig. 3 | LWIR detection near and away from EP. a**, Schematic of experimental setup for the LWIR detection measurement. **b**, Illustration of the data capturing and processing procedures from raw I/Q data to unwrapped phases and to oscillating frequencies by fitting phase. **c-f**, Measured frequency modulation (FM) waveforms and spectra when incident LWIR is modulated at (**c**) 18, (**d**) 1,920, (**e**) 6,170, and (**f**) 13,890 Hz. The incident LWIR power is 135 nW. The incident LWIR powers are calibrated to the detecting acoustic mode area of our detector. The readout bandwidth is 2 kHz in **c**, 200 kHz in **d-f**.

**Sensing performance**

We characterize the frequency responses of our nonlinear EP detector (**Figs. 4a and 4b**). When the sensor is away from EP, we observe a 3-dB bandwidth of 400 Hz and a 10-dB bandwidth of 3,500 Hz, which is limited by the thermal relaxation from the sensing resonator to the substrate (environment). Near the EP, we observe an extended 3-dB bandwidth of 3,500 Hz and 10-dB bandwidth of 7,300 Hz, which is limited by the thermal relaxation between the two coupled resonators. The thermal relaxation between two resonators (with a distance of 200 μm) is expected to be much faster than that from one resonator to the environment (for example, chip thickness of 500 μm). In other words, our EP system offers a differential detection between two resonators: the EP enhances differential perturbations between two resonators, but does not enhance any common-mode perturbations. As a results, the peak responsivity under EP condition is at about 500 Hz; when the modulating frequency is below that, the thermal relaxation between two resonators will reduce their difference in temperature. In current experimental configuration, we focus the incident LWIR on one of the resonators, it leads to a differential input, which will be largely enhanced by our EP system. Meanwhile, environmental temperature fluctuations or drifts will likely affect both resonators, and thus will not be amplified by the EP system.

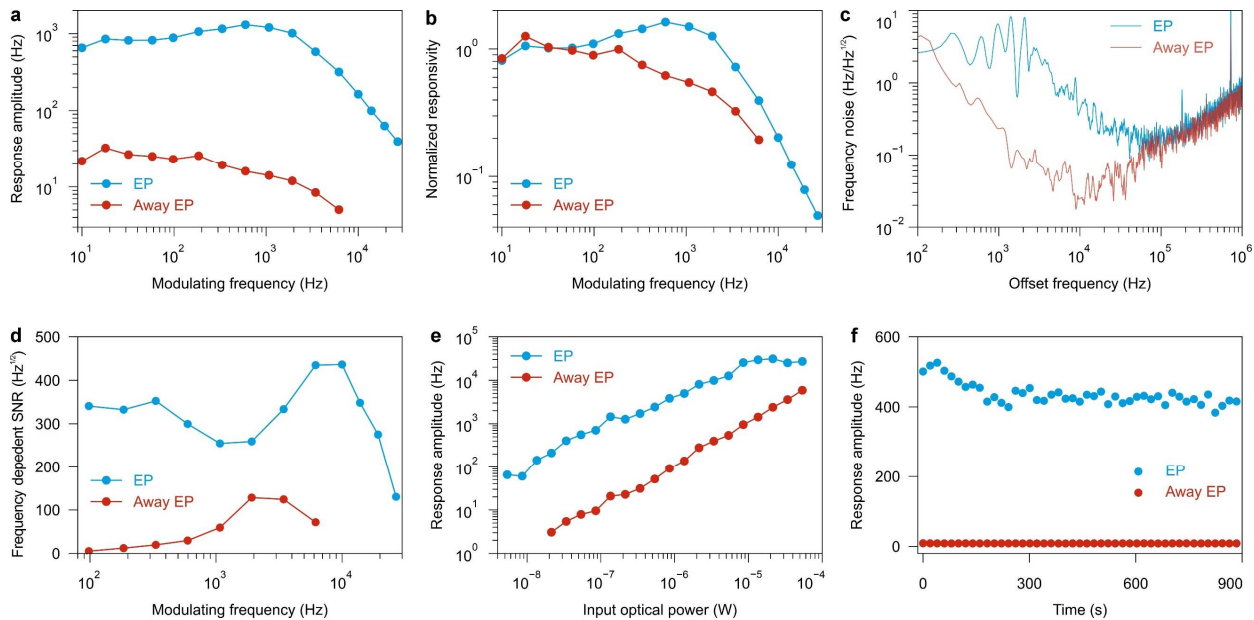

**Fig. 4 | Characterization of our LWIR detector near and away from EP. a**, Frequency response and **b**, Normalized frequency response to modulated incident LWIR showing enhanced responsivity and bandwidth when operation near EP. **c**, Measured frequency noise spectra of our detector. **d**, SNR for incident LWIR modulated at different frequencies. SNR is normalized to 1 Hz bandwidth near the signal frequency. **e**, Power dependency on the incident LWIR power. The incident LWIR power in **a** and **d** is 135 nW. Incident LWIR power is calibrated to the detecting acoustic mode area of our detector. Results in Figs. 3 and 4 are from the same device. **f**, Time stability of sensor responsivity over 15 mins without any feedback control.

We further characterize the frequency noise spectra of our detector at and away from EP (**Fig. 4c**). At the low frequencies, the measured noise is dominated by the local temperature fluctuation noises of sensor, which typical shows a 1/f behavior in frequency noise. We see that the EP system does increase the noises. Towards higher offset frequencies (>10 kHz), the noises for the case away from EP is being affected by the noise floor of readout circuits or processes, which shows a linear increase with offset frequency in the frequency noise spectrum, while it is flat in a phase noise plot (**Supplementary Fig. 3**). In contrast, as EP system significantly enhances signal, such limitation of readout circuits will happen at a much higher frequencies, here > 100 kHz, beyond the thermodynamic bandwidth of our sensors.

We define the narrowband SNR at a given frequency of interest (**Fig. 4d**) as the ratio between the response amplitude (**Fig. 4a**) and the noise spectral density (**Fig. 4c**) normalized a 1-Hz bandwidth near that frequency. Thus, either a higher responsivity or a lower noise will contribute to a better narrowband SNR. Considering this narrowband SNR, the bandwidth of our EP sensor further extends to 20 kHz. Comparing results of operations near and away from EP, the EP provides significant SNR enhancement at both low and high frequency ends, flats the SNR over the whole detecting bandwidth from 10s Hz to 20 kHz, which is highly desirable in LWIR detection.

Linearity or dynamic range of EP-based detectors are concerned for practical applications. Theoretically, at exact nonlinear EP, the change of oscillating frequency is proportional to the cube root of the input, i.e., $p_{in}^{1/3}$, and the linearity region will also be very narrow. In practice, the inevitable mismatch will reduce the responsivity but, on the other hand, increase the linear response region, leading to a trade-off between responsivity enhancement and dynamic range. At the peak responsivity, our nonlinear-EP-based sensor shows a nearly linear response up to 10 μW (**Fig. 4e**), rendering a dynamic range of 45 dB for 1-Hz measurement bandwidth (NEP of 310 pW to 10 μW). In comparison, the sensor shows linear response when operating away from EP, despite the responsivity is much smaller than that at near EP.

In addition, we characterize the stability of our nonlinear EP sensor (**Fig. 4f**). We tune the sensor to the peak responsivity, and continuously measure the output FM signal. During the first 2.5 minutes of the measurement, the responsivity degrades about 16% (likely due to stabilization of the whole experimental setup), then it remains stable for more than 10 mins without any feedback control. The long-term stability is rooted in the differential detection nature of EP sensor, as we discussed above. These results provide direct evidence that EP sensor can be sufficiently stable and robust for practical applications. We note that regular calibrations at intervals of a few minutes are common practices in most commercial thermal cameras.

The frequency-dependent narrowband SNR can be directly converted to frequency-dependent noise equivalent power (NEP) by the input power divided by the narrowband SNR (**Fig. 5**). At all tested modulation frequencies, the NEP at EP condition is significantly better than that away from the EP. With a calibrated incident LWIR power of 135 nW, our EP detector shows a minimum noise equivalent power (NEP) of 310 pW·Hz$^{-1/2}$ operating at EP at its optimal modulation frequency of 10,000 Hz, yielding a figure of merit (FoM) – NEP and time constant product (NEP·τ) – of $9.87\times10^{-3}$ pW·Hz$^{-3/2}$. In comparison, operating our detector away from EP results in an NEP of 1,086 pW·Hz$^{-1/2}$ at its optimal modulation frequency of 3,440 Hz, yielding a FoM of 0.100 pW·Hz$^{-3/2}$, which is 10 times worse than operation at EP.

We compare our detector with other mechanical-resonator-based LWIR (8-14 μm) detectors (**Fig. 5**). We see that employment of EP has pushed the performance of our detector to the state of the art. We note that the point E in **Fig. 5** represents our previous work using a single resonator with similar mechanical resonator design principles [24], indicating the EP detector with coupled resonators outperforms single resonator detector. When the coupled resonators are tuned away from the EP, the most acoustic wave energy will be confined in the active resonator, and thus the performance is expected be similar to a single-resonator detector.

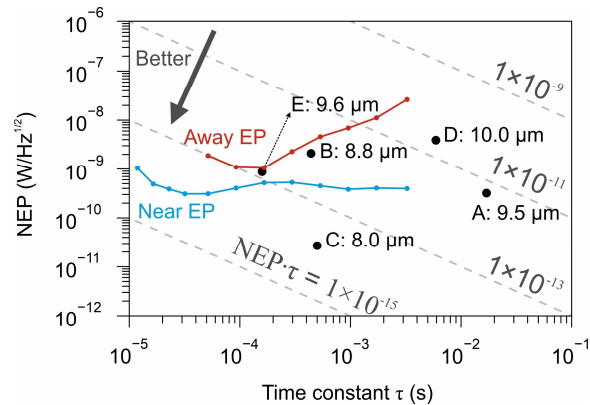

**Fig. 5 | Comparison of mechanical-resonator-based LWIR detectors.** Frequency dependent NEPs are plotted for both near EP and away from EP cases. The LWIR wavelengths of interest are 8-14 μm, only results with narrowband or peak LWIR wavelength within the range are included. NEPs and time constants from reference are A:[42], B:[43], C: [47], D:[50], E:[24].

## Discussions

From the prospect of practical applications, we emphasize that the noises of sensors are complicated, and most sensors are far above their fundamental theoretical limit of sensitivity, known as the Cramér–Rao bound (CRB) [56]. Many sensors are actually limited by the readout processes (for example, quantization noise due to limited bits of analog to digital converters), thus employing EP-based systems will significantly improve the responsivity and SNR. In addition, EP systems utilize two resonators and naturally forms a differential or balanced detector configuration, which could positively change the noise dynamics of the sensor, especially when the noises are correlated (colored), leading to non-Markovian dynamics. We emphasize that our experimental results do not contradict the recent theoretical and numerical discussions on nonlinear-EP-based sensing [28,30], where physical models with straightforward nonlinear EP systems and noises are studied. In real application scenarios, on top of theoretical (or fundamental) limits, sensing performance metrics are also determined by multiple factors including the sensing mechanics, device design, noise sources, readout processes, and thus employment of EP system can be beneficial in applications. In addition, data processing algorithms could play significant roles in the SNR of derived signal, and a well-designed algorithm could approach the Cramér–Rao bound of detector [57].

## Conclusions

We demonstrate practical benefits of employing EP system in sensors by directly comparing the sensing performance of a same device being operated at EP with that away from EP. EP system is employed in our sensors with minimal experimental overhead: while two coupled acoustic resonators are utilized in this work, the electronic circuit used in this work is the same as our previous work based on single acoustic resonator. Under practical application conditions, our sensor at EP exhibits enhancements in responsivity, bandwidth, and detected SNR and shows large dynamic ranges and good stabilities. The strategies in leveraging EP system are not limited to acoustic-wave sensors or LWIR detection, but could be applied to mechanical, optical, and electronic sensors for broad sensing applications, such as temperature, gas, particle, electromagnetic fields.

## Methods

**Design and fabrication of our detector**

We use etched grooved on 128Y-cut LN to define phononic crystals [58] and the acoustic waves are propagating in the crystal X direction. The resonant modes are formed by adjusting the pitches between etched grooves, *i.e.* periods (lattice constant) of the phononic crystal. The design of our coupled resonators is shown in **Supplementary Fig. 1**. The resonant modes have a designed resonant frequency around 1 GHz. Our single layer LWIR metamaterial absorber features a peak absorption at wavelength of 9.6 μm. The design and FDTD simulation configuration of absorber is shown in **Supplementary Fig. 3**.

The grooves are patterned by electron beam lithography using polymethyl methacrylate (PMMA) as resist, followed by inductively coupled plasma reactive ion etching (ICP-RIE) using argon gases. The IDT and metal absorbers are defined by a lift-off process including patterning by electron beam lithography with alignment to the etched patterns and a deposit of 10 nm chromium and 30 nm gold by electron beam evaporation.

**Characterization and readout of our nonlinear EP system**

As shown in Figs. 1a and 3a, the oscillation loop of our system consists of our acoustic-wave resonator, a low noise amplifier (LNA, Mini-Circuits ZX60-2534MA-S+), an electrically-controlled microwave phase shifter (Mini-Circuits JSPHS-1000+), and a microwave coupler (Mini-Circuits ZFDC-10-5-S+). We use a 6-V LiFePO$_4$ battery and a linear and low-dropout (LDO) regulator (Texas Instruments TPS7H1111EVM) to power the LNA, minimizing the noise from the power supply. An in-phase / quadrature (I/Q) demodulator (Analog Devices ADL5380), a low noise microwave generator (serving as a local oscillator (LO)) and a high-resolution oscilloscope are used to detect the oscillation signal from the coupler. The Oscilloscope captures the I/Q data from the I/Q demodulator output ports for further digital data processing.

We use a phase fitting algorithm to extract the oscillating frequency (as output signal) related to the LO frequency from the measured I/Q data. Considering measured I/Q data over a short period $t_N$, defined by the analysis bandwidth, we can write the dataset as $I_k$ and $Q_k$, for each sampling time $t_k$ where $k = 0,1,2...,N$-1. We then calculate the phase/angle $\phi_k$ of each point by treating the I/Q data as a complex number, $I_k + iQ_k$. We unwrap the phase $\phi_k$ and perform a least square fitting to the model $\hat{\phi}(t) = \omega_{sig} t + \phi_0$. The fitted slope $\omega_{sig}$ is used as the oscillating frequency. Therefore, a high sampling rate of raw I/Q data, ranging 1 to 10 Mega sample per second (MSPS) depends on the total sample time, is down sampled to the analysis bandwidth. For example, if an analysis bandwidth of 200 kHz is used in analyzing a 10 MSPS raw I/Q data, each fitting processes uses 50 data points.

**Experimental characterization of LWIR detection**

The schematic of LWIR setup is shown in **Fig. 3a**. we employ a distributed feedback laser (working wavelength at 9.6 um, QD9550C2, Thorlabs) as LWIR source. In experiments, we either use a mechanical chopper to modulate the LWIR intensity, or by directly apply the modulating signal to the laser driver. The mechanical chopper provides better performance at low modulation frequencies (e.g. less than 100 Hz), and the laser direct modulation method is used for high modulation frequency up to 20 kHz, limited by the laser driver circuits. A series of neutral density (ND) filters (#12-003 to #12-017, Edmund optics and NDIR03B, Thorlabs) are used to provide different incident LWIR power onto the detector. A low-order quarter waveplate (WPLQ05M-4500, Thorlabs) and a linear polarizer (WP25HB, Thorlabs) are used to tune the polarization of the LWIR laser. The modulated LWIR is then focused on the detector by an objective lens (#3423, Edmund optics). Similar setup was also used in our previous work [24].

**Acknowledgements**
The views and conclusions contained in this document are those of the authors and do not necessarily reflect the position or the policy of the United States Government. No official endorsement should be inferred. Approved for public release.

**Author Contributions**
L.S. conceptualized the idea. Z.X. and L.S. designed and nanofabricated the acoustic-wave devices with contributions from J.J. and T.S.. Z.X., Z.C., J.G.T., and D.W. implement and characterized the acoustic-wave EP system; L.Z. developed a low-noise power supply for the detector. Z.C., D.W., T.S., and Y.Y. designed the absorber and performed LWIR measurements. J.G.T. and Y.Z. developed and performed optical vibrometer measurements. Y.Z. analytically analyzed the system with contributions from H.L. and L.S.. L.S. developed and analyzed the numerical simulations of sensor dynamics with help from H.L..  All authors contribute to the analysis and discussions of results. L.S. drafted the text of manuscript, and Z.X. prepared the figures. All authors contribute to the revision of manuscript. L.S., Y.Y, and Y.Z. supervised the project.

# Supplementary Information

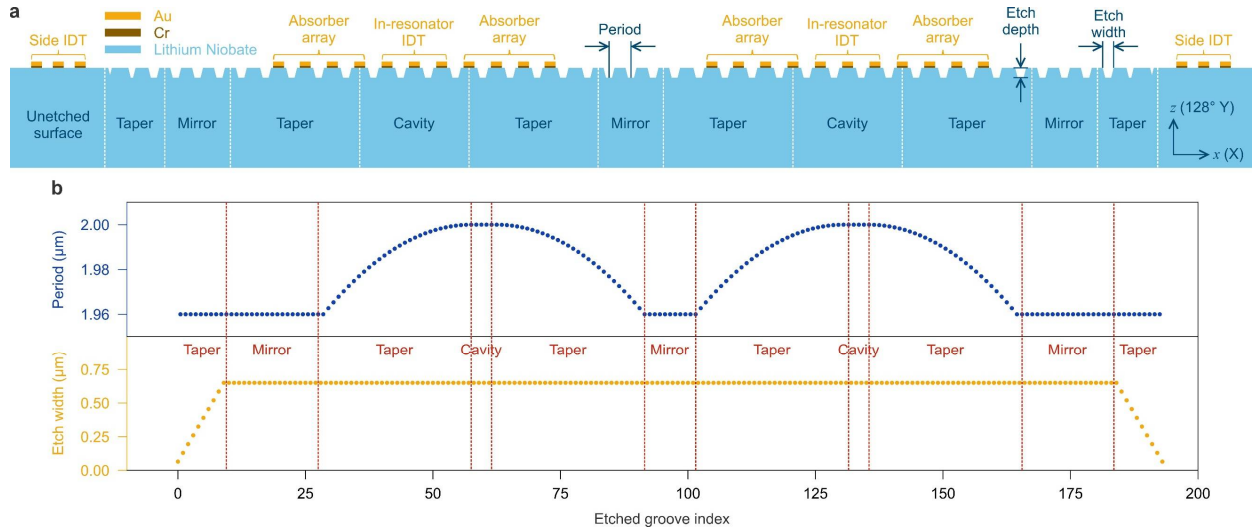

**Supplementary Figure 1 | Design of our coupled phononic crystal resonators. a**, Cross-section schematic of the resonators showing IDTs, absorbers, and different regions of phononic crystals. **b**, Period and top etch width of each phononic crystal unit cell. The number of taper and mirror cells between the two resonators determine the coupling strength.

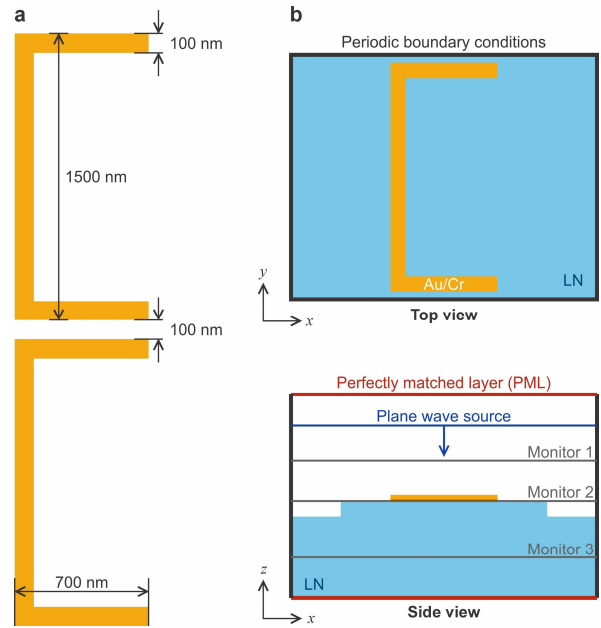

**Supplementary Figure 2 | Design and FDTD simulation of absorbers. a**, Dimensions of the 9.6-μm absorber used in this work. **b**, configuration of the FDTD simulation.

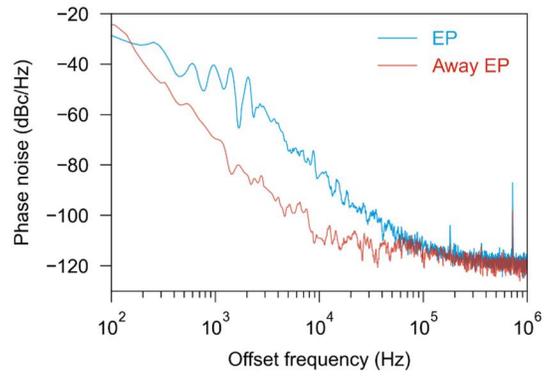

**Supplementary Figure 3 | Phase noise spectra of our detector.** This plot is based on the same measured data as that in Fig. 4c.

## Supplementary Note 1
### Steady state and responsivity of our nonlinear EP system

For the sake of conciseness, we first consider the steady state of the nonlinear-EP electromechanical systems,

$$\frac{d}{dt}\begin{pmatrix}a\\b\end{pmatrix} = -i\begin{pmatrix}\omega_a - i\frac{\gamma_a}{2} & \mu \\ \mu & \omega_b + ig_{\text{NL}}(b)e^{i\theta} - i\frac{\gamma_b}{2}\end{pmatrix}\begin{pmatrix}a\\b\end{pmatrix}. \tag{S1}$$

At steady state, the system is oscillating at a single frequency $\omega_{\text{osc}}$, and the solution can be written as $\tilde{a} = a_0 e^{-i\omega_{\text{osc}}t}$ and $\tilde{b} = b_0 e^{-i\omega_{\text{osc}}t}$, where $a_0$ and $b_0$ is the content complex amplitude.

The characteristic polynomial can be obtained as

$$\left[\omega_{\text{osc}} - \left(\omega_a - i\frac{\gamma_a}{2}\right)\right]\left[\omega_{\text{osc}} - \left(\omega_b - i\frac{\gamma_b}{2} + ig_{\text{NL}}\cos\theta - g_{\text{NL}}\sin\theta\right)\right] - \mu^2 = 0. \tag{S2}$$

From the imaginary part, the nonlinear gain is given by

$$g_{\text{NL}} = \frac{(\omega_{\text{osc}} - \omega_b)\gamma_a + (\omega_{\text{osc}} - \omega_a)\gamma_b}{2(\omega_{\text{osc}} - \omega_a)\cos\theta - \gamma_a\sin\theta}.$$

Eq. (S2) can reduce to

$$\omega_c^3 + C\omega_c + D = 0, \tag{S3}$$

$$\omega_c = (\omega_{\text{osc}} - \omega_a) - \frac{1}{3}\left(\Delta - \frac{\gamma_b}{2}\tan\theta\right),$$

$$\Delta = \omega_b - \omega_a,$$

$$C = \frac{\gamma_a^2}{4} - \mu^2 - \frac{1}{3}\left(\Delta - \frac{\gamma_b}{2}\tan\theta\right)^2,$$

$$D = -\frac{2}{27}\left(\Delta - \frac{\gamma_b}{2}\tan\theta\right)^3 - \frac{1}{3}\left(\frac{\gamma_a^2}{2} + \mu^2\right)\left(\Delta - \frac{\gamma_b}{2}\tan\theta\right) + \frac{1}{2}\gamma_a\mu^2\tan\theta.$$

We note that the depressed cubic equation in Eq. (S2) has closed-form solutions and the analytical forms above give insights for the nonlinear EP system with complex gain. The $\Delta - \frac{\gamma_b}{2}\tan\theta$ term shows the detuning between two resonators can be compensated by a proper chosen $\theta$. We can further derive the responsivity of the nonlinear EP system by differentiating Eq. S3, showing

$$\frac{\partial\omega_c}{\partial\Delta} = \frac{2}{3} + \frac{4}{3}\frac{9\mu^2 + (4\Delta - 6\omega_c)(\Delta + 3\omega_c) + \gamma_b\tan\theta\,(-4\Delta - 3\omega_c + \gamma_b\tan\theta)}{3\gamma_a^2 - 4(\Delta^2 + 3\mu^2 - 9\omega_c^2) + \gamma_b\tan\theta\,(4\Delta - \gamma_b\tan\theta)}.$$

Responsivities $R_a$ and $R_b$ of the oscillating frequency to the perturbations on the passive resonator $a$ and active resonator $b$ can be given by

$$R_a = \frac{\partial\omega_{\text{osc}}}{\partial\omega_a} = -\frac{\partial\omega_c}{\partial\Delta} + \frac{2}{3},$$

$$R_b = \frac{\partial\omega_{\text{osc}}}{\partial\omega_b} = \frac{\partial\omega_c}{\partial\Delta} + \frac{1}{3}.$$

In the case the system is largely detuned from EP point, the energy will be mainly in the active resonator $b$, and in this case $R_a \to 0$ and $R_b \to 1$. This reduces the sensing scheme to only single resonator. We note that due to the presence of nonlinear complex gain, the EP condition is not at exact $\mu = \gamma_a/2$ as in conventional EP systems.

We analytically calculate our nonlinear-EP sensor, showing agreements with experimental results and numerical simulations. The parameters used in the calculations are summarized in Supplementary Table 1. For the sake of conciseness, we shift the resonant frequency to near DC frequency; this is equivalent to analyzing the system in a proper rotating frame.

**Supplementary Table 1 | Parameters of the systems used in Supplementary Figs. 4 and 5.**

| Parameter | Unit | Value |
|---|---|---|
| $\omega_a/(2\pi)$ | Hz | 0 |
| $\omega_b/(2\pi)$ | Hz | $-0.2 \times 10^6$ |
| $\gamma_a/(2\pi)$ | Hz | $10^6$ |
| $\gamma_b/(2\pi)$ | Hz | $10^6$ |
| $2\mu/\gamma_a$ | | 0.90, 0.99, 1.1 |

We show the oscillating frequency $\omega_{\text{osc}}$, gain $g_{\text{NL}}$, and responsivity $R_b$ when tunning the gain phase $\theta$ (**Supplementary Fig. 4**). Here, we use the coupling $2\mu/\gamma_a$ = 0.90, 0.99, and 1.1 for the weak, near EP, and strong coupling cases. Under the weak coupling condition (**Supplementary Figs. 4a-4d**), there is only one real root, and the oscillating frequency $\omega_{\text{osc}}$ continues changing with the gain phase $\theta$. The gain phase $\theta$ can compensate for the mismatch in resonant frequency $\Delta$. As our sensor is oscillating, the total gain and loss of the whole system is balanced. The nonlinear gain gets maximum when the resonant frequency mismatch is compensated by the gain phase and the mode energy is most evenly distributed over the two resonators. A maximum responsivity $R_b$ of 5.6 is reached at $\theta = -12.09°$. From the perspective of the oscillating frequency, the responsivity $R_b$ shows a 3-dB bandwidth of 273 kHz. To the first order, this infers a bandwidth of linear response range; any input perturbation drives the system oscillating frequency outside this range will induce large nonlinear responsivity.

Under near nonlinear EP condition (**Supplementary Figs. 4e-4h**), a sharp change of $\omega_{\text{osc}}$ is observed when $\theta$ is tuning through the nonlinear EP. The nonlinear gain peaks at $g_{\text{NL}} \cos\theta /\gamma_a \sim 1$, which means the energy is evenly distributed over the two resonators, i.e., $|a| \cong |b|$, and thus the $g_{\text{NL}} \cos\theta$ is close to the total loss of both resonators. A much higher peak responsivity $R_b$ of 148 is observed at $\theta = -11.42°$. The responsivity $R_b$ shows a bandwidth of 42 kHz in oscillating frequency. Compared to the weak coupling case, the near-EP sensor features a much higher responsivity meanwhile a narrower linearity range.

Under strong coupling condition (**Supplementary Figs. 4i-4l**), multiple roots are possible when tunning $\theta$. From stability calculation, we learn that two modes (in blue and red) are stable, and one mode is unstable (Dashed yellow line in **Supplementary Fig. 4i**). The oscillating frequency $\omega_{\text{osc}}$ from roots of Eq. S3 is no longer monotonic but folds back onto itself, creating bifurcation. In the nonlinear gain curve (**Supplementary Fig. 4j**), it can be seen that different solutions correspond to different gain, and the system will hop to the mode with lower gain when tuning $\theta$. This mode hop leads to a forbidden range of oscillating frequency and a frequency jump when continuously tuning $\theta$, which has been experimentally observed as in **Fig. 2**. Such frequency jumps have been experimentally reported in previous works [15], but we give a clear explanation of the origin of such jumps. Enhancement in responsivity $R_b$ is observed when near the jumping frequency and is uneven at the lower and higher side of the frequency, which agrees with our experimental results. We note that system with such mode hopping may not be ideal for practical sensing applications.

We further analyze the system's dependency on the coupling $\mu$ (**Supplementary Fig. 5**). In the weak coupling region ($2\mu < \gamma_a$), there is only one real root, whereas there are three real roots in the strong coupling region. However, only two of the three roots are stable solutions but have different steady-state gain. The one with smaller gain is the solution that the system is operating. In the weak coupling limit ($\mu \ll \gamma$), the system reduces to the single resonant case, and thus the gain $g_{NL} \cos\theta \to \gamma/2$ (in this calculate we set $\gamma_b = \gamma_a$ for the sake of conciseness). In the strong coupling limit, the gain in resonator $b$ needs to compensate for the losses of both resonators, and therefore $g_{NL} \cos\theta \to \gamma$.

We plot a phase diagram of the EP system in **Supplementary Figure 6**. Under different coupling $\mu$ and gain phase $\theta$, the EP system may show (1) only one stable mode, or (2) two stable modes and one unstable mode. The stable mode requiring lower gain will be the one oscillating due to nonlinear gain competition.

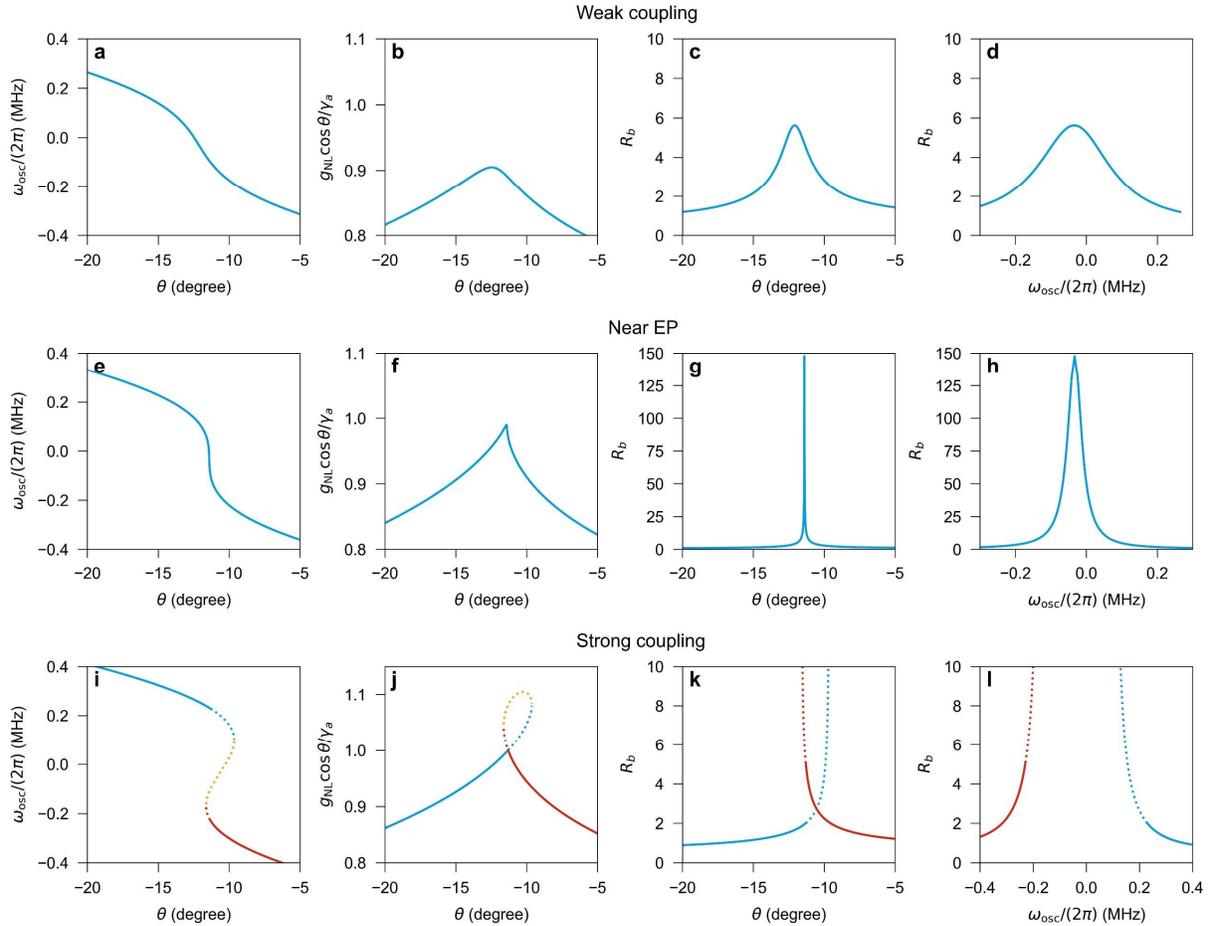

**Supplementary Figure 4 | Analytically calculated results of the nonlinear EP sensor under different coupling conditions. a-d**, weak coupling, **e-h**, near nonlinear EP, and **i-l**, strong coupling sensors. Oscillating frequency $\omega_{\text{osc}}$ (**a, e, i**), actual gain $g_{\text{NL}}\cos\theta$ (**b, f, j**), and responsivity $R_b$ (**c, g, k**), depends on the gain phase $\theta$. Responsivity $R_b$ versus the oscillating frequency $\omega_{\text{osc}}$ (**d, h, l**). In **i-l**, solid lines indicate stable and reachable solutions, orange dotted lines are unstable solutions, meanwhile, blue and red dotted lines are stable but unachievable due to nonlinear gain competing.

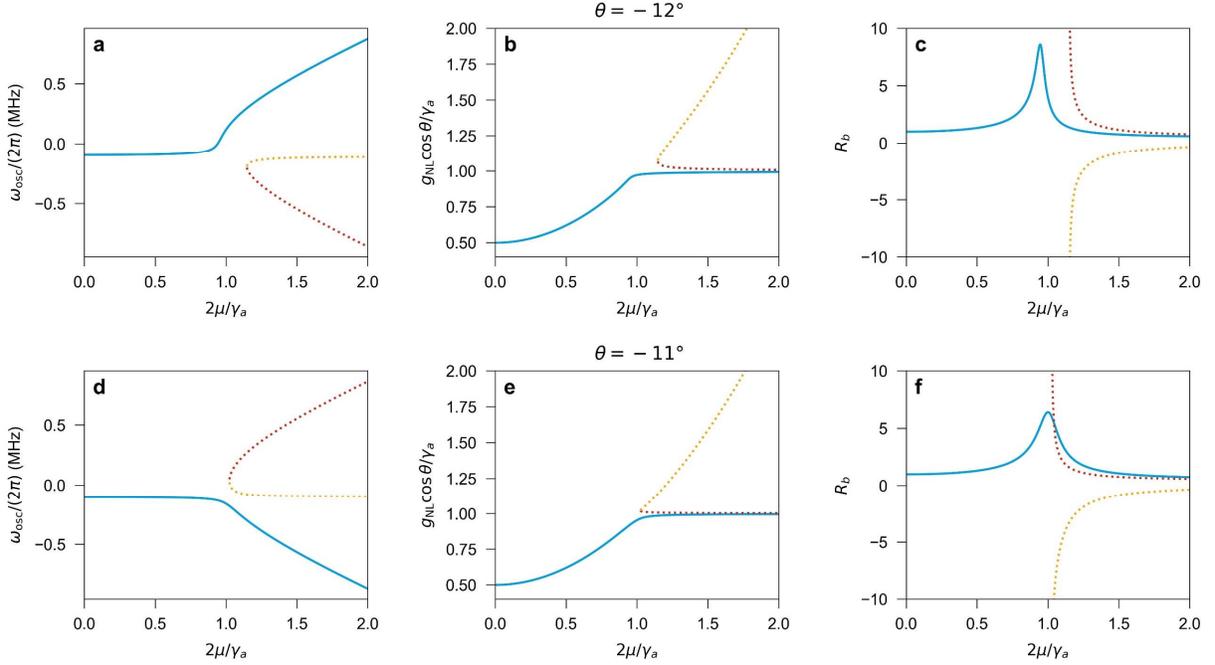

**Supplementary Figure 5 | Analytically calculated results of the nonlinear EP sensor under different gain phase conditions. a-c**, gain phase $\theta = -12°$. **d-f**, gain phase $\theta = -11°$. Oscillating frequency $\omega_{osc}$ (**a, d**), actual gain $g_{NL}\cos\theta$ (**b, e**), and responsivity $R_b$ (**c, f**), depends on the normalized coupling $2\mu/\gamma_a$. Solid lines indicate stable and reachable solutions, orange dotted lines are unstable solutions, meanwhile, red dotted lines are stable but unachievable due to nonlinear gain competing.

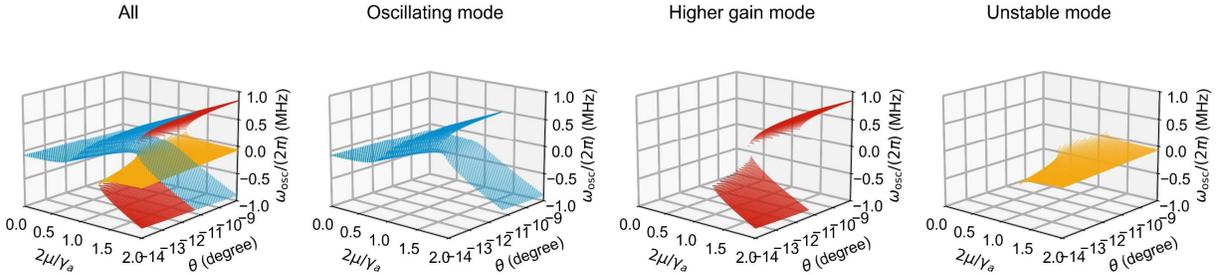

**Supplementary Figure 6 | Analytically calculated phase diagram of the nonlinear EP sensor.** Blue points indicate the stable and oscillating modes, orange points indicate the unstable mode, and red points indicate modes requiring high gain, which are stable but unachievable due to nonlinear gain competing.

## Supplementary Note 2
### Dynamics of our nonlinear-EP-based LWIR detection

The dynamics of our nonlinear-EP-based LWIR sensor includes thermodynamics and electromechanics processes, which are coupled via the thermoelastic effect, i.e., the temperature-dependent acoustic resonant frequency. The equations of motion of our sensor can be described by,

$$\frac{d\Delta T_a}{dt} = -\Delta T_a \left(\frac{1}{\tau_e} + \frac{1}{\tau_c}\right) + \frac{\Delta T_b}{\tau_c} + \zeta_a,$$

$$\frac{d\Delta T_b}{dt} = -\Delta T_b \left(\frac{1}{\tau_e} + \frac{1}{\tau_c}\right) + \frac{\Delta T_a}{\tau_c} + \frac{P_{in}}{c_p} + \zeta_b,$$

$$\frac{d}{dt}\begin{pmatrix}a\\b\end{pmatrix} = -i\begin{pmatrix} \omega_a + \eta\Delta T_a - i\frac{\gamma_a}{2} & \mu \\ \mu & \omega_b + \eta\Delta T_b + ig_{NL}(b)e^{i\theta} - i\frac{\gamma_b}{2} \end{pmatrix}\begin{pmatrix}a\\b\end{pmatrix} + \sqrt{\kappa_{b1}}\begin{pmatrix}0\\\xi_{in}\end{pmatrix}.$$

Here $\tau_e$ and $\tau_c$ are the thermal relaxation time from the resonator to the environment and the other resonator, respectively, $P_{in}$ is the absorbed power of incident LWIR, $c_p$ is the heat capacity of individual mechanical resonator, and $\kappa_{b1}$ is the external coupling rate from the side IDT. $\zeta_a$ and $\zeta_b$ are the temperature fluctuation noise of each resonator. As room temperature is much larger than the mechanical resonator frequency ($k_B T \gg \hbar\omega$, $k_B$ is the Boltzmann constant, $\omega$ is the angular frequency of the acoustic wave), the temperature fluctuation noise can be described by random variables of white Gaussian noises,

$$\overline{\zeta} = 0, \quad \overline{\zeta(t)\zeta(t')} = \frac{2\Delta T_0^2}{\tau_e}\delta(t - t'),$$

where $\Delta T_0 = \sqrt{k_B T_0^2/c_p}$ is the temperature fluctuation, $T_0 = 295$ K is the environment temperature, $\delta(t - t')$ is the Dirac delta function. The nonlinear gain can be written as,

$$g_{NL}(b) = \frac{g_0}{\sqrt{1 + \left|\frac{b}{b_{sat}}\right|^2}},$$

where $g_0 = A_0\sqrt{\kappa_{b1}\kappa_{b2}}$ is the small signal gain, $A_0$ is the net small-signal gain of the gain loop, $\kappa_{b1}$ ($\kappa_{b2}$) is the external coupling rate using the side (in-resonator) IDT, $b_{sat} = \sqrt{P_{sat}/(\hbar\omega_0\kappa_{b1})}$ is the saturation amplitude, and $P_{sat}$ is the input-referenced saturation power of LNA. For our electromechanical system, the dominated noise is the LNA-amplified thermal noise $\xi_{in}$, referenced to the input of the electromechanical system. $\xi_{in}$ can be described as a white Gaussian noise,

$$\overline{\xi_{in}(t)} = 0, \quad \overline{\xi_{in}(t)\xi_{in}(t')} = A_0^2 \frac{4 k_B T}{\hbar\omega}\delta(t - t').$$

The sensing protocol is to estimate the incident LWIR power $P_{in}$ by observing the signal out-coupled from the gain loop. Using the I/Q downconversion, we can capture the complex value (amplitude and phase) of the output signal, and the output signal can be written as

$$b_{out} = \beta\, g_{NL}(b)\, b + \xi_{read},$$

where $\beta$ is the coupling ratio of the output coupler, $\xi_{read} = \xi_{LO} + \xi_{IQ} + \xi_{scope}$ is the noise associated with the readout processing, $\xi_{LO}$ the phase noise of local oscillator used for the I/Q downconversion, $\xi_{IQ}$ is the added noise of the I/Q downconverter (based on its noise figure), and $\xi_{scope}$ is the added noise of the oscilloscope capturing the low frequency IQ signals. Depending on the actual instrument used in experiments and the interested single frequency ranges, the readout noise can be dominated by different sources.

Overall, this model well describes the dynamics and noises of our nonlinear-EP-based sensor and overall agrees with experimental observations of both response and noise behavior. We note that this model is based on several assumptions: (1) White Gaussian noises are assumed for multiple random variables, but the actual noise dynamics could be more complex, especially noises associated with micro/nanoscale structures. (2) Each resonator is described by an effective temperature and an overall mode amplitude, while the dynamics related to the spatial profile of the amplitude within the acoustic mode area is not considered.

**Supplementary Note 3**
**Numerical simulation of our nonlinear-EP-based LWIR detection**
We perform numerical simulations based on the model including thermodynamics and acoustic-wave EP system. The parameters used in the numerical simulations are summarized in **Supplementary Table 2**. Our experimental system results in a total readout noise $\xi_{read}$ of about 30 dB, which is mainly limited by the noise floor of our oscilloscope (about -140 dBm/Hz at the used signal scale). We give the numerically simulated results with (**Supplementary Fig. 7**) and without (**Supplementary Fig. 8**) this readout noise term. The difference in noise of the FM spectra can be observed between the simulation results with and without the readout noise term. Comparing the time traces in **Supplementary Fig. 7** with that in **Supplementary Fig. 8** clearly show that, away from the EP, the readout noise overwhelms the modulated output signal at 2000, 6000, and 15000 Hz, whereas the corresponding EP responses are significantly stronger. We use the same phase fitting algorithm is extract the oscillation frequency in our simulations. Overall, the numerical simulation results agree with our experimental results (**Figs. 3d-3f**).

**Supplementary Table 2 | Parameters used in numerical simulations**

| Parameter | Unit | Value |
|---|---|---|
| $\omega_a/(2\pi)$ | Hz | $-0.1 \times 10^6$ |
| $\omega_b/(2\pi)$ | Hz | $-0.1 \times 10^6$ |
| $\gamma_a/(2\pi)$ | Hz | $10^6$ |
| $\gamma_b/(2\pi)$ | Hz | $10^6$ |
| $2\mu/\gamma_a$ | | 0.99 |
| $\tau_e$ | s | $10.6 \times 10^{-4}$ |
| $\tau_c$ | s | $0.9 \times 10^{-4}$ |
| $c_p$ | J/K | $1.305 \times 10^{-9}$ |
| $g_0$ | | 20 |
| $\theta$ (EP) | degree | 0 |
| $\theta$ (Away EP) | degree | -5 |
| $\xi_{read}$, related to thermal noise floor | dB | 30 |

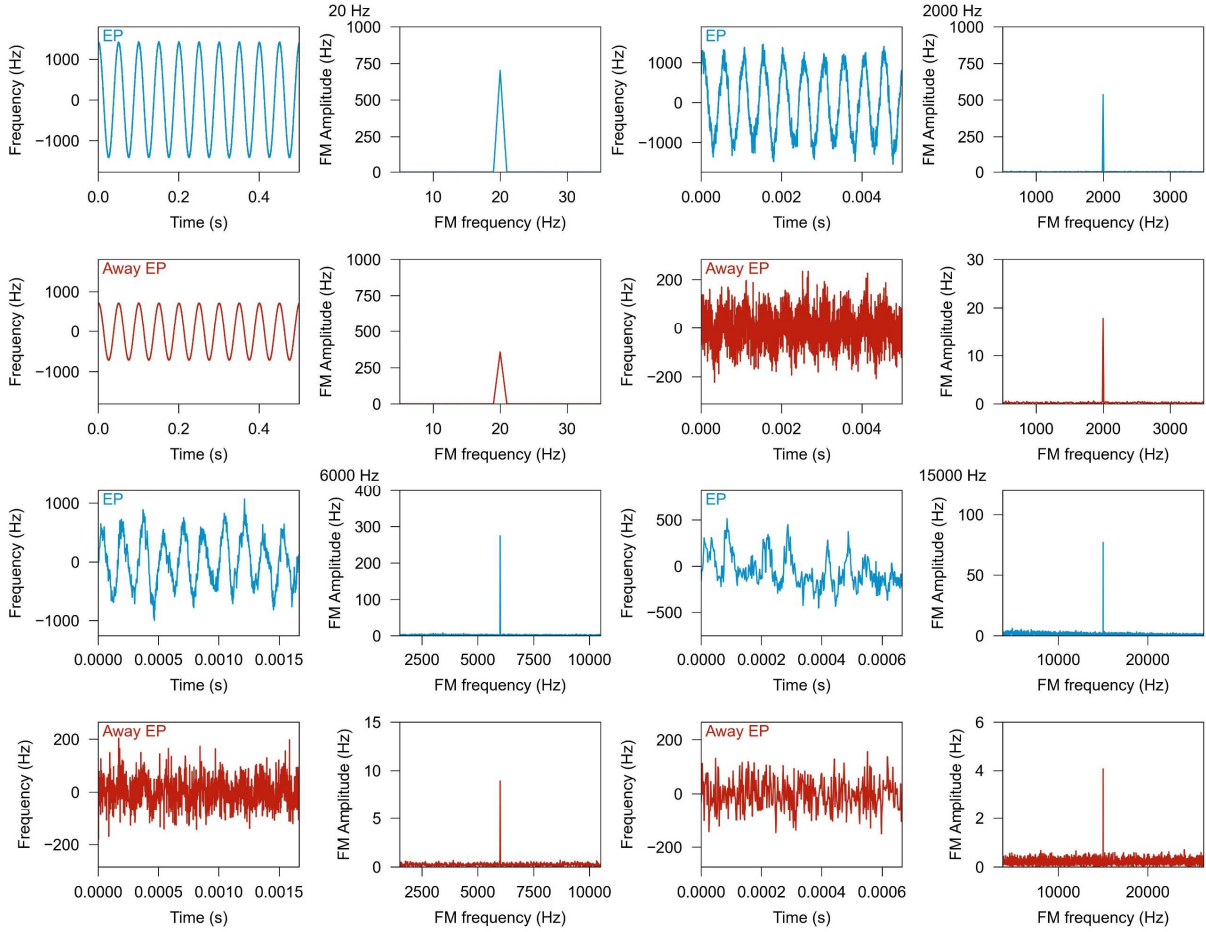

**Supplementary Figure 7 | Numerical simulation results with a readout noise of 30 dB above thermal noise limit.** The input modulation frequencies are 20, 2000, 6000, and 15000 Hz. The readout bandwidth is set to 20 kHz for 20 Hz input modulation and 200 kHz for higher input modulation frequencies.

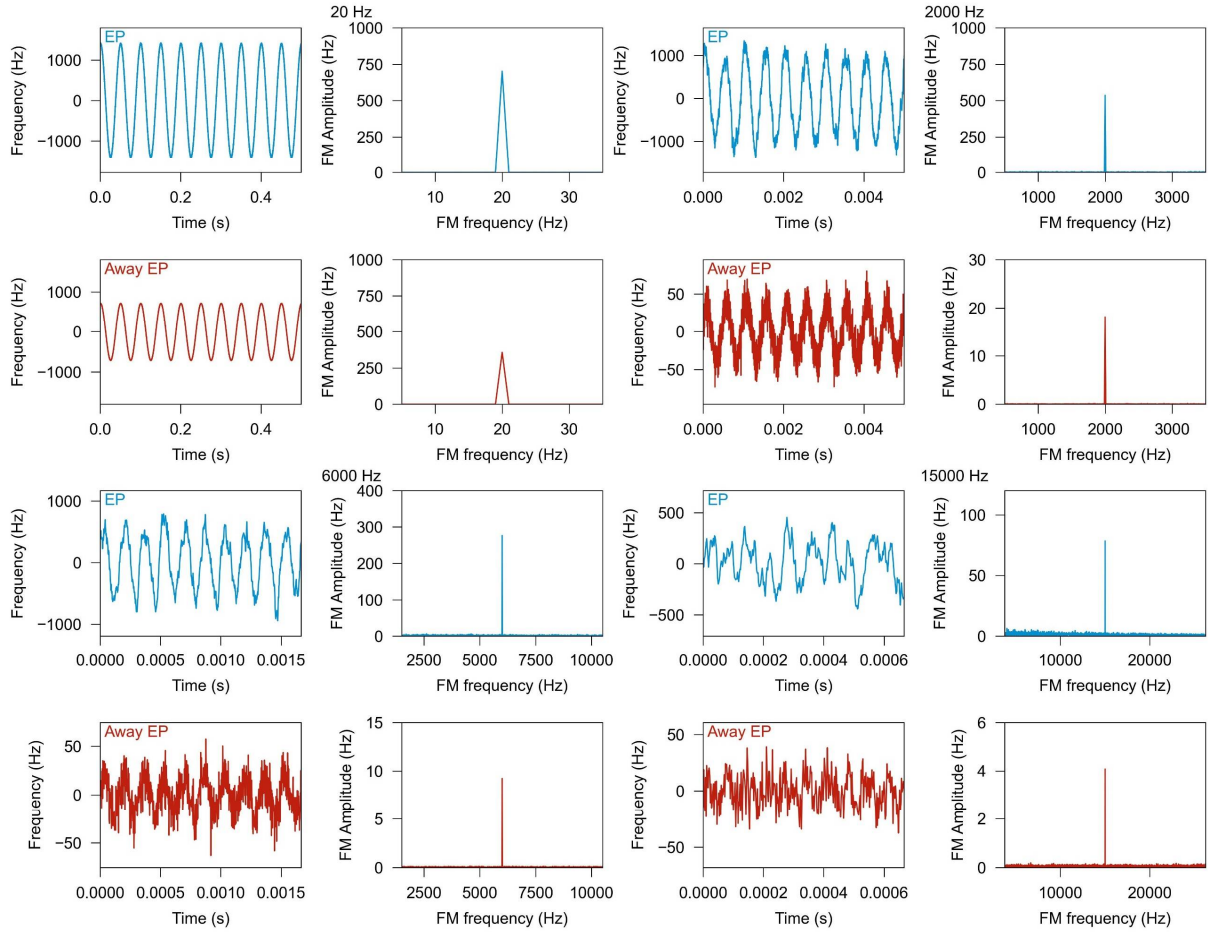

**Supplementary Figure 8 | Numerical simulation results without readout noise.** The input modulation frequencies are 20, 2000, 6000, and 15000 Hz. The readout bandwidth is set to 20 kHz for 20 Hz input modulation and 200 kHz for higher input modulation frequencies.

**Supplementary Note 4**
**Estimation of device parameters based on optical vibrometry measurements**
Accurate estimation of device parameters, including resonant frequencies, losses, coupling rates, is critical to understand how close such devices could be tuned to the EP condition by the gain phase. Conventionally, we can estimate these device parameters by fitting to the transmission spectra. When the coupling between two resonators is relatively strong, the transmission spectrum exhibits two separated peaks, and the fitting processes is reliable. However, when the coupling is weaker, the two transmission peaks will merge together, thus an accurate parameter estimation based on transmission spectrum fitting is challenging. As our work shows tuning towards EP for devices from weak to strong coupling, a more reliable parameter estimation approach is needed. Using our in-house optical vibrometry [54], we can directly measure the frequency-dependent displacement field intensity of each resonator, when the acoustic-wave is excited from one side IDT (**Supplementary Fig. 9**).

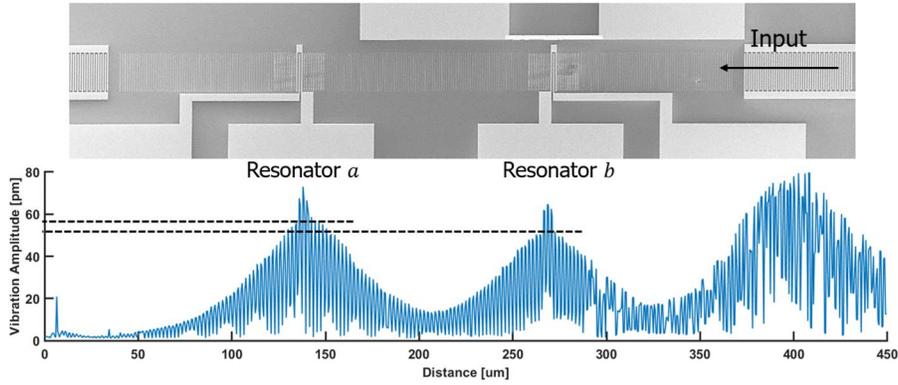

**Supplementary Figure 9** | Optical vibrometry measurement of our coupled resonators.

The expected amplitudes $a$ and $b$ of resonators can be derived by solving the coupled resonator system at its stable state. The equation of motion of two coupled resonator with excitation can be given by

$$\frac{d}{dt}\begin{pmatrix}a\\b\end{pmatrix} = -i\begin{pmatrix}\omega_a - i\frac{\gamma_a}{2} & \mu \\ \mu & \omega_b - i\frac{\gamma_b}{2}\end{pmatrix}\begin{pmatrix}a\\b\end{pmatrix} + \sqrt{\kappa_{b1}}\begin{pmatrix}0\\s_{in}\end{pmatrix}e^{-i\omega_{in}t}.$$

Here $s_{in}$ and $\omega_{in}$ is the amplitude and frequency of the excitation signal. The solution to the model is given by

$$|a_0(\omega)| = \frac{\sqrt{\kappa_{b1}}\, s_{in}\mu}{\sqrt{\left[(\omega-\omega_a)(\omega-\omega_b)-\frac{\gamma_a\gamma_b}{4}-\mu^2\right]^2 + \left[\frac{\gamma_b}{2}(\omega-\omega_a)+\frac{\gamma_a}{2}(\omega-\omega_b)\right]^2}}$$

$$|b_0(\omega)| = \frac{\sqrt{\kappa_{b1}}\, s_{in}\sqrt{(\omega-\omega_a)^2+\frac{\gamma_a^2}{4}}}{\sqrt{\left[(\omega-\omega_a)(\omega-\omega_b)-\frac{\gamma_a\gamma_b}{4}-\mu^2\right]^2 + \left[\frac{\gamma_b}{2}(\omega-\omega_a)+\frac{\gamma_a}{2}(\omega-\omega_b)\right]^2}}.$$

**Supplementary Figure 10** shows calculated amplitudes of both resonators at different excitation frequencies and for devices with different coupling. By fitting the measured amplitude to the model (**Supplementary Fig. 11**), we can estimate device parameters for all coupling conditions. **Supplementary Table 3** shows estimated parameters of five devices fabricated on the sample chip with coupling rates from weak to strong. We can see here that, by designing the numbers of periods in the taper and mirror phononic crystal structures, we can have devices with coupling condition needed by EP. A typical mismatch in $2\mu/\gamma$ is within 3%. We note that the asymmetry in the measured amplitude curves is caused by the resonant frequency mismatch between two resonators. This device parameter estimation approach is more accurate than fitting the transmission spectrum, especially for devices in the weak coupling conditions.

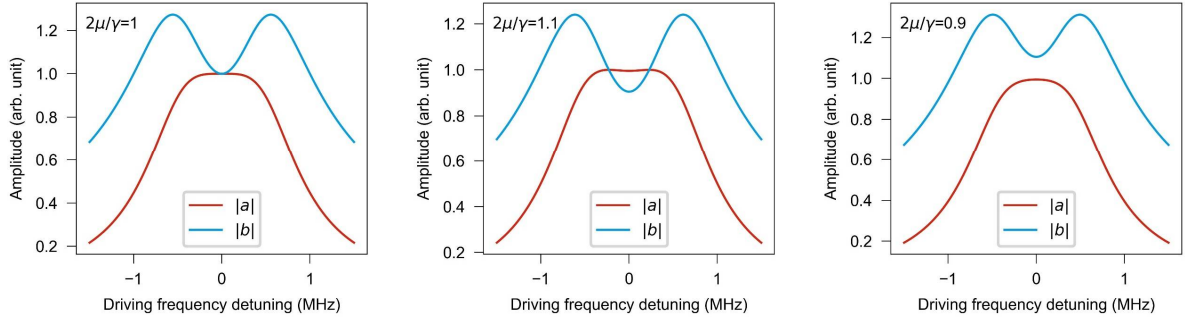

**Supplementary Figure 10 | Calculated amplitude of each resonator at different excitation frequencies $\omega_{in}$ and for devices with different coupling rate between two resonators.**

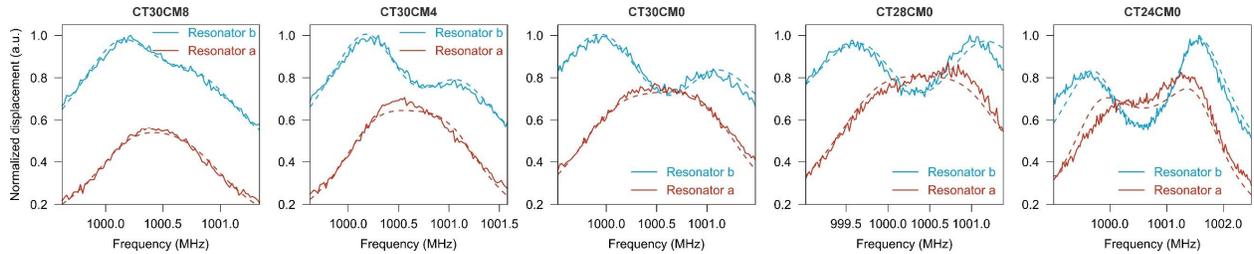

**Supplementary Figure 11 | Measured amplitude of each resonator at different excitation frequencies and for devices with different coupling rate between two resonators.** The solid lines are experimental measured amplitudes, and the dashed lines are fitted model. The device name showing on top of each plot corresponds to device parameters in **Supplementary Table 3**.

**Supplementary Table 3 | Device parameters estimated by optical vibrometry measurements.** All devices in this table are fabricated on a same chip.

| Device name | Unit | CT30CM8 | CT30CM4 | CT30CM0 | CT28CM0 | CT24CM0 |
|---|---|---|---|---|---|---|
| # Taper[1] | | 30 | 30 | 30 | 28 | 24 |
| # Mirror[1] | | 8 | 4 | 0 | 0 | 0 |
| $\mu$ | MHz | 0.333 | 0.406 | 0.493 | 0.655 | 0.935 |
| $\gamma_a/2$ | MHz | 0.55 | 0.478 | 0.482 | 0.593 | 0.816 |
| $\gamma_b/2$ | MHz | 0.386 | 0.425 | 0.62 | 0.801 | 0.338 |
| $2\mu/\gamma_a$ | | **0.605** | **0.849** | **1.023** | **1.105** | **1.146** |
| $\omega_a/(2\pi)$ | MHz | 1000.545 | 1000.727 | 1000.626 | 1000.397 | 1000.552 |
| $\omega_b/(2\pi)$ | MHz | 1000.395 | 1000.518 | 1000.402 | 1000.411 | 1000.717 |
| $\Delta/(2\pi)$ | MHz | -0.151 | -0.209 | -0.224 | 0.014 | 0.165 |

[1]Number of periods of taper and mirror structures between the two resonators, which define the coupling strength between two resonators.